\documentclass{elsart1p}
\usepackage{graphicx}
\usepackage{amssymb}
\newenvironment{myitemize} {
  \begin{list}{--}
      {
	\setlength{\leftmargin} {4mm}
	\setlength{\parsep}     {0pt}
	\setlength{\itemsep}    {0mm}
	\setlength{\topsep}     {\itemsep}
	\setlength{\partopsep}  {0pt}
	\setlength{\parskip}    {0pt}
	}} {
  \end{list}}

\newcommand{\eVq}{\ensuremath{\text{eV}^2}}

\usepackage{amsmath,amssymb}
\usepackage{graphicx}
\usepackage{rotating}
\usepackage{array}
\begin{document}
\begin{frontmatter}
% use the corauthref command within \author for corresponding author
% footnotes;
\title{Neutrino Physics}
\author{M.C. Gonzalez-Garcia}
\address{Instituci\'o Catalana de Recerca i Estudis Avan\c{c}ats (ICREA) \&
  Departament d'Estructura i Constituents de la Mat\`eria, Universitat
  de Barcelona, Diagonal 647, E-08028, Spain\\
and\\
Y.I.T.P., SUNY at Stony Brook, Stony Brook, NY 11794-3840, USA}

\begin{abstract}
In this talk I will review our present knowledge on neutrino masses
and mixing trying to emphasize  the most
direct implications and challenges of these results.
\end{abstract}
%
%\begin{keyword}
%\PACS
%\end{keyword}
\end{frontmatter}
\section{Introduction: The New Minimal Standard Model}
The SM is a gauge theory based on the gauge symmetry $SU(3)_{\rm C}\times SU(2)_{\rm
L}\times U(1)_{\rm Y}$ spontaneously broken to $SU(3)_{\rm C}\times
U(1)_{\rm EM}$ by the the vacuum expectation value of 
a Higgs doublet field $\phi$.  As I will explain in Sec.\ref{impli},
with the matter contents required for describing the observed 
particle interactions,  
the SM predicts that neutrinos are {\sl strictly}  massless
and there is neither mixing nor CP violation in the
leptonic sector.

We now know that this picture cannot be correct. Over several years
we have accumulated important experimental evidence that neutrinos
are massive particles and there is mixing in the leptonic sector.
In particular we have learned that:
\begin{myitemize}
\item Solar $\nu_e's$ convert to $\nu_{\mu}$ or $\nu_\tau$ with 
confidence level (CL) of more than 7$\sigma$~\cite{solar}.
\item KamLAND find that reactor $\overline{\nu}_e$ disappear over distances
of about 180 km and they observe a distortion of their energy spectrum.
Altogether their evidence has more than 3$\sigma$ CL~\cite{kamland}.
\item The evidence of atmospheric (ATM) $\nu_\mu$  disappearing is now at
$> 15 \sigma$, most likely converting to $\nu_\tau$~\cite{skatm}.
\item  K2K observe the disappearance of accelerator $\nu_\mu$'s at 
distance of 250 km and find a distortion of their energy spectrum 
with a CL of 2.5--4 $\sigma$~\cite{k2k}.
\item  MINOS observes the disappearance of accelerator $\nu_\mu$'s at 
distance of 735 km and find a distortion of their energy spectrum 
with a CL of $\sim$ 5 $\sigma$~\cite{minos}.
\item  LSND found evidence for 
$\overline{\nu_\mu}\rightarrow\overline{\nu_e}$. This evidence has not
been confirmed by any other experiment so far and it is being tested
by MiniBooNE \cite{minib}. 
\end{myitemize}
These results imply that neutrinos are massive and the Standard Model
has to be extended at least to include neutrino masses. 
This minimal extension is what I call 
{\sl The New Minimal Standard Model}. 

In the New Minimal Standard Model flavour is mixed in the CC
interactions of the leptons, and a leptonic mixing matrix appears
analogous to the CKM matrix for the quarks.  However, the discussion of
leptonic mixing is complicated by two factors. First the number of
massive neutrinos ($n$) is unknown, since there are no constraints on
the number of right-handed, SM-singlet, neutrinos. Second, since
neutrinos carry neither color nor electromagnetic charge, they could
be Majorana fermions. As a consequence the number of new parameters in
the model depends on the number of massive neutrino states and on
whether they are Dirac or Majorana particles.

In general, if we denote the neutrino mass eigenstates by $\nu_i$,
$i=1,2,\ldots,n$, and the charged lepton mass eigenstates 
by $l_i=(e,\mu,\tau)$, in the mass basis, leptonic CC interactions 
are given by
\begin{equation}
-{\cal L}_{\rm CC}={g\over\sqrt{2}}\, \overline{{l_i}_L} 
\, \gamma^\mu \, U_{ij}\,  \nu_j \; W_\mu^+ +{\rm h.c.}.
\label{CClepmas}  
\end{equation} 
Here $U$ is a $3\times n$ matrix 
$U_{ij}=P_{\ell,ii}\, {V^\ell_{ik}}^\dagger \, V^\nu_{kj}\, (P_{\nu,jj})$
where $V^\ell$ ($3\times 3$) and $V^\nu$ ($n\times n$) are the diagonalizing
matrix of the charged leptons and neutrino mass matrix respectively
${V^\ell}^\dagger M_\ell M_\ell^\dagger V^\ell=
{\rm diag}(m_e^2,m_\mu^2,m_\tau^2)$ and 
${V^\nu}^\dagger M_\nu^\dagger M_\nu V^\nu={\rm diag}(m_1^2,m_2^2,m_3^2,\dots,
m_n^2)$.
 
$P_\ell$ is a diagonal $3\times3$ phase matrix, that is conventionally
used to reduce by three the number of phases in $U$.
$P_\nu$ is a diagonal matrix with additional arbitrary phases (chosen
to reduce the number of phases in $U$) only for Dirac states. 
For Majorana neutrinos, this matrix is simply a unit matrix, 
the reason being that if one rotates a Majorana neutrino 
by a phase, this phase will appear in its
mass term which will no longer be real.
Thus, the number of phases that can be absorbed by redefining the
mass eigenstates depends on whether the neutrinos are Dirac or Majorana 
particles. In particular, if there are only three Majorana (Dirac) neutrinos, 
$U$ is a $3\times 3$ matrix analogous to the CKM matrix for the
quarks
but due to the Majorana (Dirac) nature of the neutrinos it depends on 
six (four) independent parameters: three mixing angles and three (one) phases.

A consequence of the presence of the leptonic mixing is the
possibility of flavour oscillations of the neutrinos.  Neutrino
oscillations appear because of the misalignment between 
the interaction neutrino eigenstates and the propagation eigenstates (
which  for propagation in vacuum  are the mass eigenstates). 
Thus a neutrino of energy $E$ produced in a CC
interaction with a charged lepton $l_\alpha$ can be detected via a CC
interaction with a charged lepton $l_\beta$ with a probability which
presents an oscillatory behaviour, with oscillation lengths
given by the phase difference between the different propagation eigenstates
-- which in the ultrarelativistic limit is 
$L_{0,ij}^{\rm osc}=\frac{4 \pi E}{\Delta m_{ij}^2}$ -- and amplitude
that is proportional to elements in the mixing matrix.

It follows that neutrino oscillations are only sensitive to mass squared
differences and do not give us information on the absolute value
of the masses. Also the Majorana phases do not affect oscillations because
total lepton number is conserved in the process. 
Experimental information on absolute neutrino
masses can be obtained from Tritium $\beta$ decay experiments
 \cite{tritium} and from
its effect on the cosmic microwave background radiation and large
structure formation data. If neutrinos are Majorana
particles their mass and also additional phases can be determined in
$\nu$-less $\beta\beta$ decay experiments \cite{bb}.

Besides the flavour vacuum oscillations, described above,  
further flavour dependent effects occur 
when neutrinos travel through regions of dense matter.
This is so,  because they can
undergo forward scattering with the particles in the medium and  these
interactions are, in general, flavour dependent and as a consequence
the oscillation pattern is modified. However  the flavour
transition probability still
depends only on the mass squared differences and it is independent of
the Majorana phases.

The neutrino experiments described above have measured some
non-vanishing $P_{\alpha\beta}$ and from these measurements we have
inferred all the positive evidence that we have on the non-vanishing
values of neutrino masses and mixing as described below.
\begin{figure}[t]
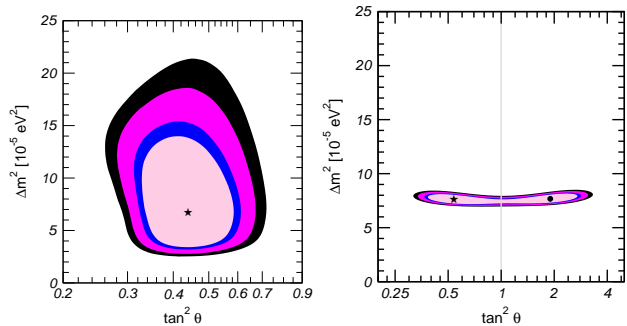

\begin{center}
\includegraphics[width=0.3\textwidth]{2nu.solar.eps} 
\includegraphics[width=0.3\textwidth]{2nu.kamland.eps} 
\end{center}
\caption{
Allowed regions for 2-$\nu$ oscillations of 
solar $\nu_e$ (left) and  KamLAND $\bar\nu_e$ (right).
The different contours correspond to the allowed regions 
at 90\%, 95\%, 99\% and $3\sigma$ CL.}
\label{fig:kland}
\end{figure}
\begin{figure}[t]
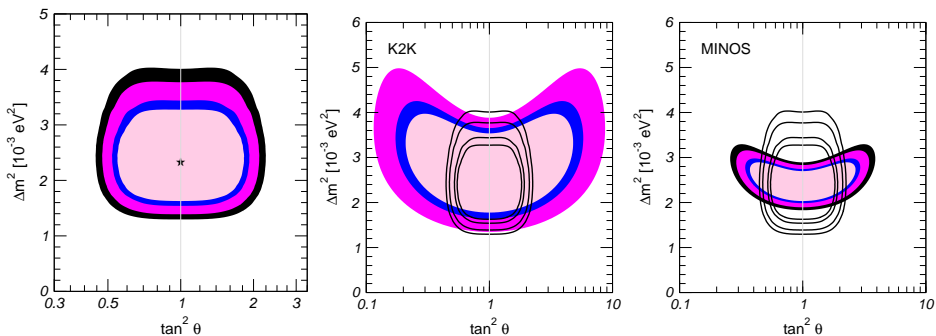

\includegraphics[width=0.3\textwidth]{2nu.atmos.eps} 
\includegraphics[width=0.3\textwidth]{2nu.k2k.eps} 
\includegraphics[width=0.3\textwidth]{2nu.minos.eps} 
\caption{
      Allowed regions from the analysis of
      ATM data (left), K2K (central) and MINOS (right). 
  The different contours correspond to at
      90\%, 95\%, 99\% and $3\sigma$ CL.}
\label{fig:atm2fam} 
\end{figure}
\section{The Parameters of the NMSM: 3$\nu$ Analysis}
I describe here the present determination of the parameters of the
model from the analysis of the data from 
solar, KamLAND, ATM and K2K experiments and ignore the LSND
evidence as it was not confirmed by MiniBoone.

In Fig.~\ref{fig:kland} I show the results from a recent
analysis~\cite{myreview2} of KamLAND $\overline{\nu}_e$ 
disappearance data and solar $\nu_e$ data.
The main features of these results are: 
\begin{myitemize}
  \item In the analysis of solar data, only the formerly called
large mixing angle solution (LMA) is allowed with maximal mixing 
rejected at more than $5\sigma$. 
This is so since the release of the SNO salt-data
    (SNOII) in Sep 2003.   
\item The analysis of the KamLAND data 
determines  more precisely $\Delta m^2_{\odot}$ and it 
already makes an impact on 
the lower bound of the corresponding mixing angle, whereas the upper bound 
is dominated by solar data, most importantly by the CC/NC solar neutrino 
rates measured by SNO. 
\item There is a mismatch between the best fit mixing angle 
and $\Delta m^2$ from solar and KamLAND at the $\sim$ 1.5$\sigma$ level.
\end{myitemize}
In Fig.\ref{fig:atm2fam} I show the results of an updated analysis
of the ATM, K2K and MINOS data (updated from \cite{myreview2}).
As seen in the figure, the determination of the corresponding 
$\Delta m^2_{atm}$ is dominated
by the data from the MINOS experiment. 
The measurement of the mixing angle is
still largely dominated by atmospheric neutrino data from
Super-Kamiokande with a best fit point close to maximal mixing. 

From the previous partical analysis, it is clear that 
the minimum joint description of ATM, LBL, solar and reactor
data requires that all the three known neutrinos take part in the
oscillations.  The mixing parameters are encoded in the $3
\times 3$ lepton mixing matrix which can be conveniently parametrized
in the standard form 
\begin{equation}
U={\small\begin{pmatrix}
    1&0&0 \cr
    0& {c_{23}} & {s_{23}} \cr
    0& -{s_{23}}& {c_{23}}\cr\end{pmatrix}
    \begin{pmatrix}
    {c_{13}} & 0 & {s_{13}}e^{i {\delta}}\cr
    0&1&0\cr 
    -{ s_{13}}e^{-i {\delta}} & 0  & {c_{13}}\cr
\end{pmatrix} %\nonumber \\
%    &&
\begin{pmatrix}
    c_{21} & {s_{12}}&0\cr
    -{s_{12}}& {c_{12}}&0\cr
    0&0&1\cr\end{pmatrix}}
\end{equation}
where $c_{ij} \equiv \cos\theta_{ij}$ and $s_{ij} \equiv
\sin\theta_{ij}$. The angles $\theta_{ij}$ can be taken without loss of
generality to lie in the first quadrant, $\theta_{ij} \in [0,\pi/2]$.  

There are two possible mass orderings, which we denote as {\sl Normal}
and  {\sl Inverted}.  In the normal scheme $m_1<m_2<m_3$ while in 
the inverted one $m_3< m_1<m_2$. 

In total the 3-$\nu$ oscillation analysis involves six
parameters: 2 mass differences (one of which can be positive or negative), 
3 mixing angles, and the CP phase. Generic 3-$\nu$ oscillation effects
include: (i) coupled oscillations with two different wavelengths;
(ii) CP violating effects; (iii) difference between Normal and Inverted schemes.The strength of these effects is controlled by the values of the ratio
of mass differences $\Delta m^2_{21}/|\Delta m_{31}^2|$, by the
mixing angle $\theta_{13}$ and by the CP phase $\delta$. 

From the previous 2$\nu$ analysis we see that 
$\Delta m^2_\odot = \Delta m^2_{21} \ll 
|\Delta m_{31}^2|\simeq|\Delta m_{32}^2|=\Delta m^2_{\rm atm}$.
As a consequence the dominant oscillations in the joint 3-$\nu$ analysis 
behave as follows:
\begin{myitemize}
  \item for solar and KamLAND neutrinos, the oscillations with the
    $\Delta m^2_{31}$-driven oscillation length are completely 
    averaged and the survival probability takes the form:
    \begin{equation}
	P^{3\nu}_{ee}
	=\sin^4\theta_{13}+ \cos^4\theta_{13}P^{2\nu}_{ee} 
	\label{eq:p3}
    \end{equation}
    where in the Sun $P^{2\nu}_{ee}$ is obtained with the modified sun
    density $N_{e}\rightarrow \cos^2\theta_{13} N_e$. So the analyses
    of solar data constrain three of the six parameters: $\Delta
    m^2_{21}, \theta_{12}$ and $\theta_{13}$. 
  \item For ATM and LBL neutrinos, the $\Delta m^2_{21}$-driven 
   wavelength is very
    long and the corresponding oscillating phase is almost negligible. As a
    consequence, the ATM and LBL data analysis mostly restricts $\Delta
    m^2_{31}\simeq \Delta m^2_{32}$, $\theta_{23}$ and $\theta_{13}$,
    the latter being the most relevant parameter common to both 
   solar+KamLAND and
    ATM+LBL neutrino oscillations and which may potentially allow
    for some mutual influence. The effect of $\theta_{13}$ is to add a
    $\nu_\mu\rightarrow\nu_e$ contribution to the ATM and LBL oscillations.
  \item In reactor experiments at short and intermediate baselines, 
   in particular  at CHOOZ, the $\Delta m^2_{21}$-driven wavelength is 
    unobservable 
    and the relevant oscillation wavelength is determined by $\Delta
    m^2_{31}$ and its amplitude by $\theta_{13}$. 
\item The CP phase is basically unobservable although there is 
some marginal sensitivity in the present ATM neutrino analysis
~\cite{myreview2}. 
\item Normal versus Inverted orderings could be discriminated due to
matter effects in the Earth for ATM neutrinos. However, this
effect is controlled by the mixing angle $\theta_{13}$. Presently 
all data favour small $\theta_{13}$. 
Consequently, the difference between Normal and
Inverted orderings is too small to be statistically meaningful in the 
present analysis. 
\item  
The previously mentioned mismatch between the best fit angle $\theta_{12}$
as determined   by solar versus KamLAND experiments, can be resolved by 
a non vanishing $\theta_{13}$ \cite{fogli,valle,michele}. The 
exact CL for the non-zero value depends on the details of the analysis.
A careful comparison of the results can be found in 
Ref.\cite{michele} in which the best fit $\theta_{13}$ is finally found 
to be compatible  with zero at the $0.9\sigma$ level.
\end{myitemize}

Altogether the derived ranges for the $\Delta m^2$'s at $1\sigma$ ($3\sigma$)
are:
\begin{equation} \begin{aligned}
    \label{eq:3nuranges}
    \Delta m^2_{21}
    &= 7.67 \,_{-0.21}^{+0.22} \,\left(_{-0.61}^{+0.67}\right)
    \times 10^{-5}~\eVq \,,
    \\
    \Delta m^2_{31} &=
    \begin{cases}
	-2.39 \pm 0.12 \,\left(_{-0.40}^{+0.37}\right)
	\times 10^{-3}~\eVq & \text{(inverted hierarchy)} \,,
	\\[1mm]
	+2.49 \pm 0.12 \,\left(_{-0.36}^{+0.39}\right)
	\times 10^{-3}~\eVq & \text{(normal hierarchy)} \,,
    \end{cases}
\end{aligned}
\end{equation}
while our present knowledge of the
moduli of the mixing matrix $U$ yields:
\begin{equation}
    |U|_{3\sigma} =
    \begin{pmatrix} 
	0.77 \to 0.86 & ~\quad~ 0.50 \to 0.63 & ~\quad~ 0.00 \to 0.22 \\
	0.22 \to 0.56 & ~\quad~ 0.44 \to 0.73 & ~\quad~ 0.57 \to 0.80 \\
	0.21 \to 0.55 & ~\quad~ 0.40 \to 0.71 & ~\quad~ 0.59 \to 0.82
    \end{pmatrix} \,.
\end{equation}

\section{Implications}
\label{impli}
\subsection{The Need of New Physics}
The SM is based on the gauge symmetry
$SU(3)_{\rm C}\times SU(2)_{\rm L}\times U(1)_{\rm Y}$  
spontaneously broken to $SU(3)_{\rm C}\times U(1)_{\rm EM}$ by the 
the vacuum expectations value (VEV), $v$, of the a Higgs doublet 
field $\phi$.
The SM contains three fermion generations which reside in chiral 
representations of the gauge group. Right-handed fields are included 
for charged fermions as they are needed to build the electromagnetic
and strong currents.  
As a consequence no right-handed neutrino is included in the
model since neutrinos are neutral and colorless. 

In the SM, fermion masses arise from the Yukawa interactions which 
couple the right-handed fermion singlets to the left-handed fermion 
doublets and the Higgs doublet. After spontaneous electroweak symmetry 
breaking (EWSB) these interactions lead to charged fermion masses 
but leave the neutrinos massless. No Yukawa interaction can be written that
would give a tree level mass to the neutrino because no right-handed 
neutrino field exists in the model. 

Furthermore, within the SM $G_{\rm SM}^{\rm global}=U(1)_B\times U(1)_e\times
U(1)_\mu\times U(1)_\tau$ is an accidental global symmetry.  Here
$U(1)_B$ is the baryon number symmetry, and $U(1)_{e,\mu,\tau}$ are
the three lepton flavor symmetries. In principle one can think of a 
neutrino mass term  built with the particle content of the SM 
--$ {Y^\nu_{ij}\over v} 
\left(\bar L_{Li} \tilde \phi\right)
\left( \phi^+ L^C_{Lj}\right) +{\rm h.c.}$, where $L_i$ are the lepton 
doublets -- as induced  perturbatively at higher order or by 
non-perturbative effects.
However  such term  would violate the
$U(1)_L$ subgroup of $G_{\rm SM}^{\rm global}$ and therefore cannot be
induced by loop corrections.  Also, it cannot be induced by
non-perturbative corrections because the $U(1)_{B-L}$ subgroup of
$G_{\rm SM}^{\rm global}$ is non-anomalous.

It follows that the SM predicts that neutrinos are precisely
massless. Consequently, there is neither mixing nor CP violation in the
leptonic sector. Thus the simplest and most straightforward 
lesson of the evidence
for neutrino masses is also the most striking one: {\sl there is
New Physics (NP) beyond the SM}. This is the first
experimental result that is inconsistent with the SM.
\subsection{The Scale of New Physics} 
There are many good reasons to think that the SM is not a
complete picture of Nature and some new physics (NP) is expected to
appear at higher energies. In this case the SM is an effective 
low energy theory valid  up to the scale $\Lambda_{\rm NP}$ 
which characterizes the NP. In this approach, 
the gauge group, the fermionic spectrum, and the pattern of 
spontaneous symmetry breaking
are still valid ingredients to describe Nature at 
energies $E\ll\Lambda_{\rm NP}$. 
The difference between the SM as a complete description of Nature
and as a low energy effective theory is that in the latter case we must
consider also non-renormalizable (dim$>4$) terms whose effect 
will be suppressed by powers $1/\Lambda_{\rm NP}^{\rm dim-4}$. 
In this approach the largest effects at low energy are expected 
to come from dim$=5$ operators 

There is a single set of dimension-five terms that is made of 
SM fields and is consistent with the gauge symmetry  given by
\begin{equation}
{\cal O}_5={Z^\nu_{ij}\over \Lambda_{\rm NP}}
\left(\bar L_{Li} \tilde \phi\right)
\left( \phi^+ L^C_{Lj}\right)+{\rm h.c.},
\label{dimfiv}  
\end{equation}
which violates  total lepton number by two units and leads, upon 
EWSB, to neutrino masses:
\begin{equation}
(M_\nu)_{ij}={Z^\nu_{ij}}{v^2\over\Lambda_{\rm NP}}.
\label{nrmass}  
\end{equation}
This is a Majorana mass term. 

Eq.~(\ref{nrmass}) arises in a generic extension of the
SM which means that neutrino masses are very likely to appear
if there is NP. Furthermore from Eq.~(\ref{nrmass}) 
we find that the scale of neutrino masses 
is suppressed by $v/\Lambda_{\rm NP}$  when compared to the scale 
of charged fermion masses providing an explanation not only for 
the existence of neutrino masses but also for their smallness. 
Finally, Eq.~(\ref{nrmass}) breaks not only total lepton number but also
the lepton flavor symmetry. 
Thus we should expect lepton mixing and CP violation.

Given the relation (\ref{nrmass}), $m_\nu\sim v^2/\Lambda_{\rm NP}$, it is
straightforward to use measured neutrino masses to estimate the scale
of NP that is relevant to their generation. In particular,
if there is no quasi-degeneracy in the neutrino masses, the heaviest
of the active neutrino masses can be estimated,
$m_h=m_3\sim\sqrt{\Delta m^2_{\rm 31}}\approx 0.05\ {\rm eV}$
(in the case of inverted  hierarchy the implied scale is 
$m_h=m_2\sim\sqrt{|\Delta m^2_{\rm 31}|}\approx 0.05\ {\rm eV}$). 
It follows that the scale in the non-renormalizable term (\ref{dimfiv})
is given by
\begin{equation}
\Lambda_{\rm NP}\sim v^2/m_h\approx10^{15}\ {\rm GeV}.
\label{estlnp}
\end{equation}
We should clarify two points regarding Eq.~(\ref{estlnp}):

1. There could be some level of degeneracy between the neutrino masses that 
are relevant to the atmospheric neutrino oscillations. In such a case 
Eq.~(\ref{estlnp}) becomes an upper bound on
the scale of NP.

2. It could be that the $Z_{ij}$ couplings of Eq.~(\ref{dimfiv}) are much
smaller than one. In such a case, again, Eq.~(\ref{estlnp}) becomes an upper 
bound on the scale of NP. On the other hand, in models of approximate flavor 
symmetries,
there are relations between the structures of the charged lepton and neutrino
mass matrices that give quite generically  $Z_{33}\geq m_\tau^2/v^2\sim
10^{-4}$. We conclude that the likely range of $\Lambda_{\rm NP}$
that is implied by the atmospheric neutrino results is given by
\begin{equation}
10^{11}\ {\rm GeV}\leq\Lambda_{\rm NP}\leq 10^{15}\ {\rm GeV}.
\label{ranlnp}
\end{equation}

The estimates (\ref{estlnp}) and (\ref{ranlnp}) are very exciting.
First, the upper bound on the scale of NP is well below the
Planck scale. This means that there is a new scale in Nature which
is intermediate between the two known scales, the Planck scale
$m_{\rm Pl}\sim10^{19}$ GeV and the electroweak breaking scale,
$v\sim10^2$ GeV. 
Second, the scale $\Lambda_{\rm NP}\sim10^{15}$ GeV is intriguingly close
to the scale of gauge coupling unification. 

Of course, neutrinos could be conventional Dirac particles.
In the minimum realization of this possibility, one must still extend
the SM to add right-handed neutrinos and {\sl impose} the conservation 
of total lepton number (since in the presence of right-handed neutrinos
total lepton number is not an accidental symmetry) to prevent
the right-handed neutrinos from acquiring a singlet Majorana mass term.
In this scenario, neutrinos could acquire a mass like any other fermion
of the Standard Model and no NP scale would be implied. 
We would be left in the darkness on the reason of the smallness 
of the neutrino mass.  

\subsection{The Challenge of Reconstructing the New Physics}

In order to illustrate what I mean with this  title 
I will focus on the what is probably the best known 
scenario that leads to (\ref{dimfiv}): the {\it see-saw mechanism}.

In what it is also called Type-I see-saw, ~\cite{seesaw}, one
assumes the existence of heavy sterile neutrinos $N_i$. Such fermions
have SM gauge invariant bare mass terms and Yukawa interactions :
\begin{equation}
-{\cal L}_{NP}=\frac{1}{2} {M_N}_{ij}\overline{N^c_i}N_j+
Y^\nu_{ij}\overline{L_{Li}}\tilde\phi N_j +{\rm h.c.}.
\label{sinint}  
\end{equation}
The resulting mass matrix 
in the basis $\left(\nu_{Li}, N_j\right)^T$  
has the following form:
\begin{equation}
M_\nu=\begin{pmatrix}
0&Y^\nu{v\over\sqrt2}\cr (Y^\nu)^T{v\over\sqrt2}&M_N
\end{pmatrix}
\label{fumama}  
\end{equation}
If the eigenvalues of $M_N$ are all well above the electroweak breaking scale
$v$, then the diagonalization of $M_\nu$ leads to three light mass eigenstates
and an effective low energy interaction  of the form (\ref{dimfiv}). 
In particular, the scale
$\Lambda_{\rm NP}$ is identified with the mass scale of the heavy sterile
neutrinos, that is the typical scale of the eigenvalues of $M_N$.
Two well-known examples of extensions of the SM that lead to
a see-saw mechanism for neutrino masses are SO(10) GUTs
and left-right symmetry.

Another form of new physics which also 
leads to a see-saw mechanism is  the Type-II see-saw\cite{TypeII}.
In this case, no additional neutrino states are included in the 
theory but in order to construct a gauge invariant neutrino mass term 
involving only left-handed neutrinos  the Higgs sector of
the Standard Model is extended to include besides the doublet $\phi$, 
an $SU(2)_L$ scalar triplet $\Delta \sim (1,3,1)$. We can write the 
triplet in the matrix representation as
\begin{equation}
\Delta = \left( \begin{array} {cc}
 \Delta^0 &  -\Delta^{+}/\sqrt{2} \\
 -\Delta^{+}/\sqrt{2}  & -\Delta^{++}
\end{array} \right).
\end{equation}
The neutrino mass term arises from the Lagrangian:
\begin{equation}
{\cal L}_{NP} =  - {f_\nu}_{ij} \ \bar L^C_{L,i} \Delta \ L_{Lj}-
V(\phi,\Delta)
\end{equation}
where the scalar potential, besides the usual mass and self-coupling terms 
for the doublet contains additional pieces such as a triple mass term
$M_\Delta$ and a double-triple mixing  term $\mu$
(which breaks $L$ explicitly)
\begin{equation}
V(\phi,\Delta)_{NP}
=M_{\Delta}^2 \ {\rm Tr} (\Delta^\dagger \Delta)
\ + \ \left( \mu\ \tilde\phi^T\ \Delta\ \tilde\phi \ + \ h.c.\right) \  \nonumber. \\
\end{equation}
The potential minimization leads to a vev for the triplet
\begin{equation}
v_{\Delta} 
= \frac{\mu \ v^2}{ \sqrt{2} \ M_{\Delta}^2}.
\label{eq:vdel}
\end{equation}
which induces a Majorana mass for the three neutrinos
\begin{equation}
M_\nu=\sqrt{2} f_\nu \, v_\Delta \, .
\label{eq:mnut2}
\end{equation}
So if $M_{\Delta}^2\gg \mu\, v$, then  $v_{\Delta}\ll v$ which 
gives an explanation to the smallness of the neutrino mass. 
In this case the
scale $\Lambda_{\rm NP}=M_{\Delta}^2/\mu$, and a characteristic 
setting would be for $f_\nu \approx 1$ and
$\mu \sim M_{\Delta} \approx 10^{14-15}$ GeV.

One may notice that even in these particularly simple forms of NP,
${\cal L}_{\rm NP}$ contains very different high--energy particle
contents as well as 18 parameters for Type-I 
and $>12$ for Type-II which we would need to know
in order to fully determine the dynamics of the NP. However the
effective low energy operator ${\cal O}_5$ contains only 9 parameters
which is everything we can measure at the low energy experiments.
This simple example illustrates the limitation of the
``bottom-up'' approach in deriving model independent implications of
the presently observed neutrino masses and mixing. 
This is the challenge. 

Alternatively one can go  ``top-down'' by studying the low energy 
effective neutrino masses and mixing induced by specific high
energy models. Unfortunately the number of possible models is 
overwhelming and impossible to review in this talk. 
 
The bottom line  of this discussion is that in order to advance 
further in the understanding of the dynamics underlying neutrino masses, 
we need more  (and more precise) data. Furthermore synergy among different 
types of observations are probably going to be fundamental in this
advance. 
In this respect I will finish by discussing two possible 
consequences of neutrino mass models which have deserved special attention 
in the last years.

\subsection{Signatures at LHC}
Ideally in order to directly test the dynamics underlying the neutrino 
mass generation  one needs to  observe the associated new states. 
For instance, in the examples above we would like  to produce and study 
the new heavy states responsible for the see-saw mechanism, either the heavy
neutral leptons or the triplet scalars. 
As discussed above, most generically the characteristic
mass scales of the new states are very large, rendering
the new states experimentally inaccessible in the foreseeable future.
However, one can envision scenarios in which this may not be necessarely 
the case.  

As an example let's take the Type II see-saw introduced in
the previous section.   The key ingredient that makes this model 
testable at LHC is to  assume a very small doublet-triplet mixing 
\begin{equation}
\mu\ll M_\Delta
\label{eq:smalldt}
\end{equation} 
so the Higgs triplet is heavy,
typically $M_{\Delta}^2 > v^2/2$, but not much heavier than this bound.  
Once the neutral component  in the triplet gets the vev, $v_{\Delta}$
as in Eq.~(\ref{eq:vdel}), the neutrinos acquire a Majorana mass 
as described above, but because of the smallness of $\mu$ 
one can have small neutrino masses even
with $M_{\Delta}$ as light as 
\begin{equation}
M_{\Delta}\sim 110\; {\rm GeV}
\end{equation}
without conflicting with any existing data. 

After EWSB, there are 7 physical
massive Higgs bosons left in the spectrum. One is a SM-like 
doublet while the other 6 are triplet-like: two neutral ($H_2$ and $A$), 
two single charged $H^\pm$ and two double charged $H^{\pm\pm}$
with 
$M_{H_2}\simeq M_{A} \simeq M_{H^{\pm}} \simeq M_{H^{++}}=M_{\Delta}$.
Thus all these states are within reach of the LHC.

In the physical basis for the fermions 
the Yukawa interactions of the single and double charged scalars can be written as
\begin{eqnarray}
-{\cal L}=Y^+_{ij} \ \bar \nu^C_{Li} 
\ e_{Lj} \ \ H^+ \ + \ \bar Y^{++}_{ij}\  \bar e_{Li}^C 
\ e_{Lj} \ H^{++} ,
\end{eqnarray}
where
\begin{eqnarray}
Y^+  =  \cos \theta_+ \ \frac{m_\nu^{diag}}{v_{\Delta}} \ U_{LEP}^\dagger, 
\;\;\;\;\;\;\;
Y^{++} =  U_{LEP}^* \ \frac{m_{\nu}^{diag}}{\sqrt{2} \ v_{\Delta}} 
\ U_{LEP}^{\dagger}=f_\nu \ . 
\label{gamma}
\end{eqnarray}
Thus in this scenario the values of the couplings $Y^+$ and $Y^{++}$ are
determined by the spectrum and mixing angles for the active neutrinos.
Therefore, by observing lepton-number violating decays 
of the Higgs bosons, 
$H^{++} \to e_i^+ e_j^+$ and $H^{+} \to e_i^+ \bar{\nu}\ (e_i=e,\mu,\tau)$ 
one can obtain information about neutrino masses 
and mixings and in particular it is possible to determine the
neutrino mass spectrum.

\subsection{Leptogenesis}
Even if its characteristic NP scale is as high as
estimated in Eq.(\ref{ranlnp}), neutrino mass models can 
lead to important visible consequences. 
In particular they may help us to explain 
the origin of the cosmic matter-antimatter asymmetry, 
via leptogenesis~\cite{lepto}.   

From the detail cosmological data from CMB and BBN  we know
that there is only a tiny asymmetry in the baryon number,
$n_B/n_\gamma \approx 5 \times 10^{-10}$.  
Leptogenesis~\cite{lepto} is the possible origin of
such a small asymmetry related to  neutrino physics.
\begin{figure}
\begin{center}
\includegraphics[width=\columnwidth]{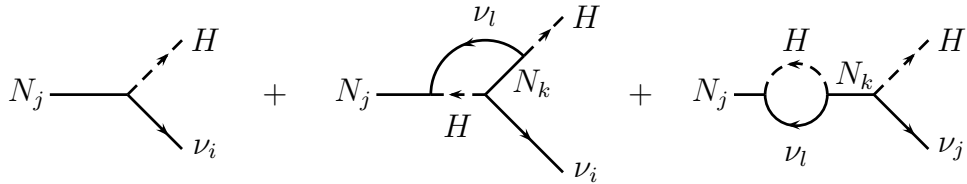}
\vglue -0.3cm
\caption{The tree-level and one-loop diagrams of
right-handed neutrino decay into leptons and Higgs.}  
\label{fig:lepto}
\end{center}
\end{figure}
In a possible realization of leptogenesis, 
$L \neq 0$ is generated  in  the 
Early Universe by the decay of one of the heavy  right-handed neutrinos
of the the see-saw mechanism, with a direct CP
violation. Due to the
interference between the tree-level and one-loop diagrams shown 
in Fig.~\ref{fig:lepto}
the decay rates of the right-handed neutrino into leptons
and anti-leptons are different. In order to generate a
lepton asymmetry the decay must be out of equilibrium 
($\Gamma_{\nu_R}\ll$ Universe expansion rate).
Sphaleron processes transform the lepton asymmetry in 
baryon asymmetry and below the electroweak phase transition 
a net baryon asymmetry is generated
$\Delta B\simeq -\frac{\Delta \rm L}{2}$ 
(the exact coefficient relating $\Delta B $
to ${\Delta \rm L}$ is model dependent.)

In general, the details are model dependent and much work has
been done to explain the observed asymmetry in realizations 
which are able to accommodate  the neutrino oscillation data. 
In its minimal implementation a right-handed neutrino 
of about $10^{10}$~GeV  can account for the cosmic baryon asymmetry 
from its out-of-equilibrium decay~\cite{leptoreview}. 

{\bf Acknowledgments}:
This work is supported by National Science Foundation
grant PHY-0354776 and  by Spanish Grants 
FPA-2007-66665-C02-01 and CSD2008-00037.

\end{document}